\documentclass[%
 reprint, superscriptaddress,
 amsmath,amssymb,
]{revtex4-2}

\usepackage[english]{babel}
\usepackage{graphicx}
\usepackage{siunitx}
\usepackage{hyperref}
\usepackage{bm}% bold math

\DeclareSIUnit\electronmass{\text{\ensuremath{m}}_{\mathrm{e}}}
\DeclareMathAlphabet\mathbfcal{OMS}{cmsy}{b}{n}
\newcommand{\appref}[1]{Appendix~\ref{#1}}

\begin{document}

\preprint{APS/123-QED}

\title{\textbf{Mechanical enhancement of quantum oscillations}}

\author{Maximilian Daschner}
\email{md867@cam.ac.uk}
\email{maximilian.daschner@lmu.de}
\affiliation{Cavendish Laboratory, University of Cambridge, Cambridge CB3 OHE, United Kingdom}
\affiliation{Fakultät für Physik, Ludwig-Maximilians-Universität, München, Germany}

\author{Ivan Kokanović}
\email{kivan@phy.hr}
\affiliation{Department of Physics, Faculty of Science, University of Zagreb, Zagreb, Croatia}
\affiliation{Cavendish Laboratory, University of Cambridge, Cambridge CB3 OHE, United Kingdom}

\author{F. Malte Grosche}
\email{fmg12@cam.ac.uk}
\affiliation{Cavendish Laboratory, University of Cambridge, Cambridge CB3 OHE, United Kingdom}

%\date{\today}

\begin{abstract}
We investigate quantum oscillation measurements in the Dirac nodal-line semimetal TaNiTe$_5$ which exhibit a strongly enhanced amplitude in the magnetoresistance. We show that mechanical properties of the measurement setup in combination with de Haas - van Alphen oscillations in the magnetic torque can cause this enhancement in the measured resistance, without involvement of any topological properties in this material. To support the empirical data, a numerical model is provided, showing good agreement.
\end{abstract}

\maketitle

\section{Introduction}
Weyl and Dirac semimetals are materials whose band structure hosts a linear band crossing in the Brillouin zone, usually close to the Fermi energy. These materials have recently gained interest due to their topological properties \cite{lv2021experimental,armitage2018weyl,yan2017topological,xiao2010berry,ong2021experimental,wang2017quantum,hasan2010colloquium}. Novel experimental signatures, some of which include the appearance of Fermi arcs \cite{xu2015discovery} and the presence of the chiral anomaly \cite{huang2015observation,xiong2015evidence}, were first discovered in these materials.

When exposed to strong magnetic fields, metals in general exhibit so-called quantum oscillations that can be probed \cite{shoenberg2009magnetic} in various physical properties such as the magnetisation, known as the de Haas - van Alphen (dHvA) effect, or the magnetoresistance (MR), known as the Shubnikov - de Haas (SdH) effect. These oscillations allow to map out the Fermi surface of a material and provide additional information about the effective masses and the nature of electron scattering. What motivates the field of probing quantum oscillations in Dirac semimetals is that under certain circumstances, they obtain a $\pi$ phase shift which can be directly related to the presence of non-trivial Berry curvature around the Dirac nodal points \cite{mikitik1999manifestation}.

Among the various material candidates for Dirac semimetals, the family TaXTe$_5$ (X = Ni, Pd, Pt) \cite{xu2020anisotropic,wang2022coexistence,liimatta1989synthesis,hu2022transport,ma2023quasi,ye2022anisotropic,huang2023magnetic,li2022coexistence,chen2021three,hao2021multiple,zhou2024origin,lu2022topologically,jiao2020topological,mar1991synthesis,xiao2022dirac,jiao2021anisotropic} falls into the category of Dirac nodal-line semimetals where the linear band crossing is not limited to a single point but extends along a line in the Brillouin zone. In this work we focus on one representative of this family, namely TaNiTe$_5$, and discuss the strongly amplified quantum oscillations in the magnetoresistance presented in a recent publication by the authors \cite{daschner2024probing}. This strong enhancement suggests the presence of a novel topological signature in Dirac nodal-line semimetals. However, a theoretical model based on classical mechanics and electrodynamics can fully account for this experimental result. A numerical study is also provided and shows good agreement with the empirical data.

\section{Methods}
Single crystals of TaNiTe$_5$ were grown with the self-flux growth method. Elements of Ta, Ni, and Te were mixed together with a molar ratio of Ta:Ni:Te = 1:1:10 inside a glove box filled with purified Argon gas. After filling the mixture into a quartz tube and sealing the tube in vacuum, it was then heated to over \SI{973}{\kelvin} over the course of four days and then cooled to \SI{773}{\kelvin} within a week. Similar to its sister compounds, TaNiTe$_5$ grows in shiny needle shaped layered single crystals.

\begin{figure}[ht]
  \centering
  \includegraphics[width=\columnwidth]{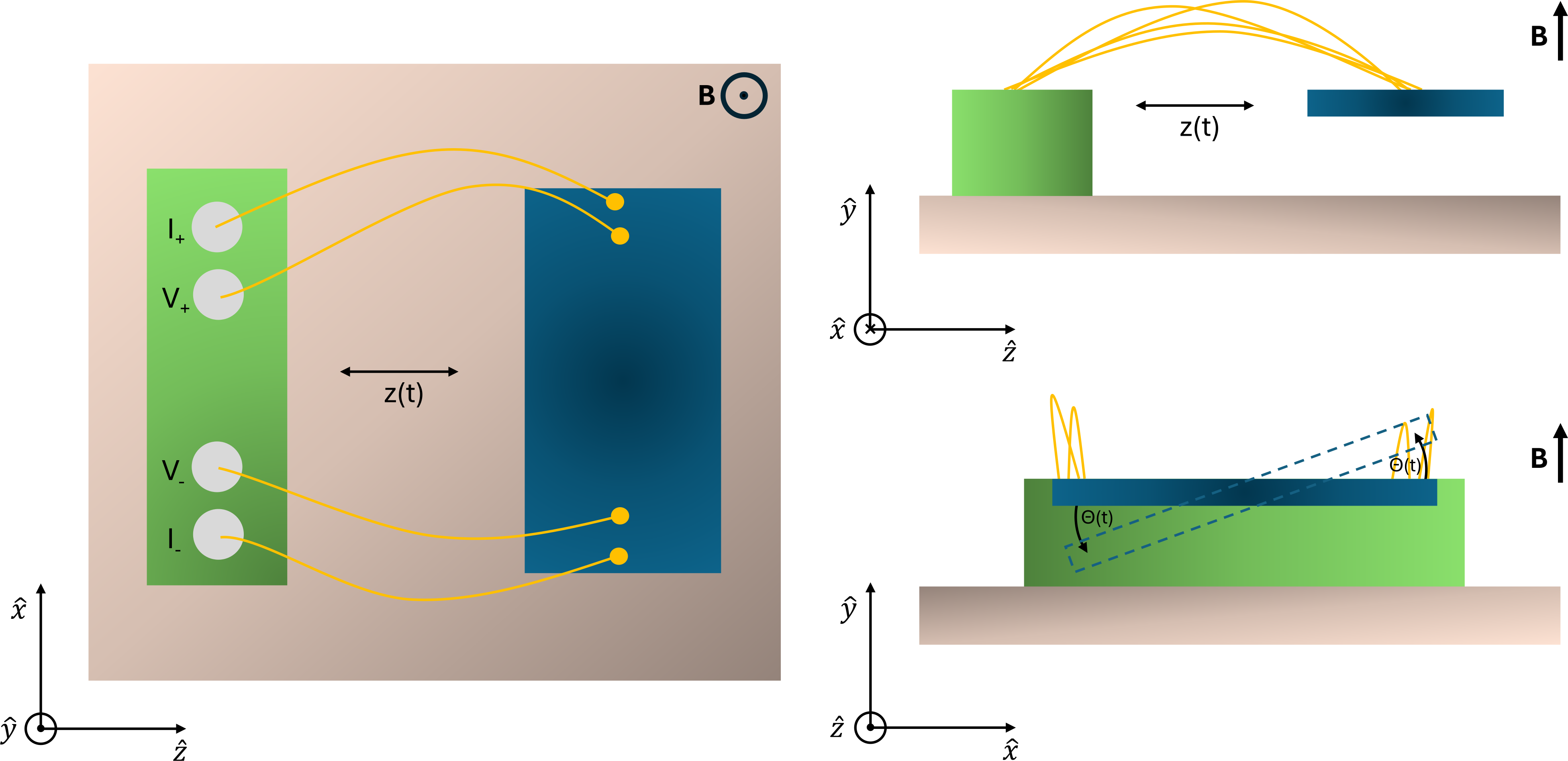}
  \caption[Typical setup of a four-point resistivity measurement.]{Typical setup of a four-point resistivity measurement of a sample (blue). The gold wires connect the sample with contact pads on a circuit board (green). Voltage (V) and current (I) contact pads from which electronic measurements can be conducted are drawn in grey. An image of a real setup as used in the \SI{9}{\tesla}-PPMS is shown in \appref{TaNiTe5_on_puck}. The crystallographic axes \textit{a}, \textit{b} and \textit{c} of TaNiTe$_5$ are aligned with $\hat{\textbf{x}}$, $\hat{\textbf{y}}$ and $\hat{\textbf{z}}$, respectively.}
  \label{TaNiTe5_measurement_sketch_1}
\end{figure}

Measurements shown here were performed in a Quantum Design Physical Property Measurement System (PPMS) with a \SI{14}{\tesla} magnet (referred to as the \SI{14}{\tesla}-PPMS), and a PPMS with a \SI{9}{\tesla} magnet (referred to as the \SI{9}{\tesla}-PPMS). A typical four-point measurement inside the cryostat is sketched in \autoref{TaNiTe5_measurement_sketch_1}. The sample is held up by gold wires and thus floating, such that it can freely move if exposed to a force. The only relevant degrees of freedom here are given by the lateral movement in \textbf{z(t)} and a rotational movement in the $\hat{\textbf{x}}$-$\hat{\textbf{y}}$ plane by an angle $\theta(t)$. All transport measurement were performed with a SR830 lock-in amplifier.

\section{Experimental Results}

\begin{figure}[t]
  \centering
  \includegraphics[width=\columnwidth]{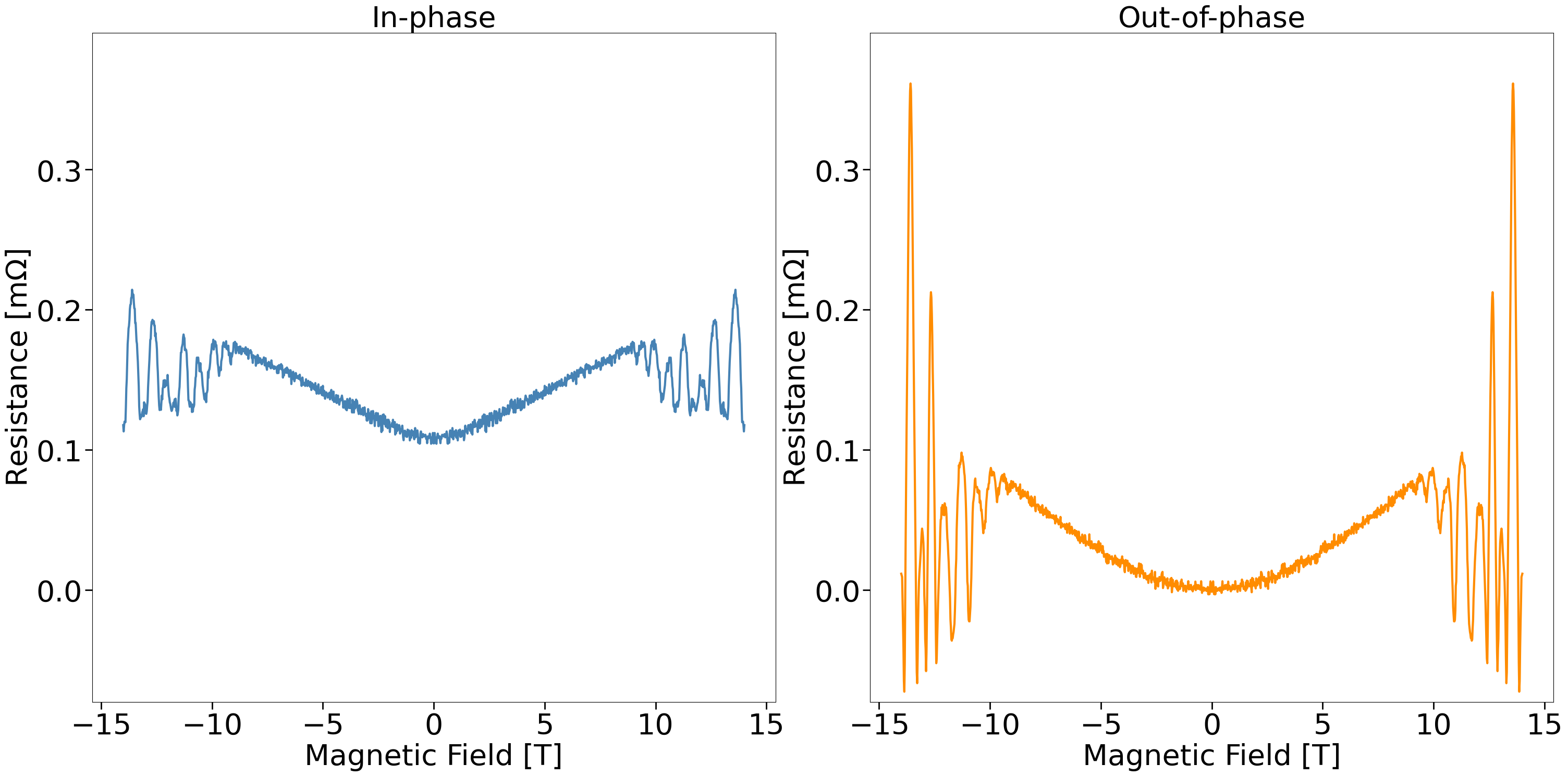}
  \caption[Magnetoresistance measurement of a TaNiTe$_5$ sample up to \SI{14}{\tesla}.]{Magnetoresistance measurement of a TaNiTe$_5$ sample where $I\parallel\hat{\mathbf{x}}$ and $B\parallel\hat{\mathbf{y}}$ (in \autoref{TaNiTe5_measurement_sketch_1}). Data is split into the in-phase (left) and out-of-phase (right) components. This data was measured in a \SI{14}{\tesla}-PPMS at \SI{1.8}{\kelvin}.}
  \label{TaNiTe5_magic_osc}
\end{figure}

\autoref{TaNiTe5_magic_osc} shows a measurement of SdH oscillations in a TaNiTe$_5$ sample, which was performed in a \SI{14}{\tesla}-PPMS with $I$ parallel to the crystallographic \textit{a} axis ($\hat{\mathbf{x}}$ in \autoref{TaNiTe5_measurement_sketch_1}) and $B$ parallel to the crystallographic \textit{b} axis ($\hat{\mathbf{y}}$ in \autoref{TaNiTe5_measurement_sketch_1}). Voltage contacts were not aligned in a straight line and hence a Hall component showed up in the measurement data. The longitudinal resistance $R(B)$ data was corrected for the Hall component by symmetrising via
\begin{equation}
	R = \frac{R(B)+R(-B)}{2}
\end{equation}

\begin{figure}[t]
  \centering
  \includegraphics[width=\columnwidth]{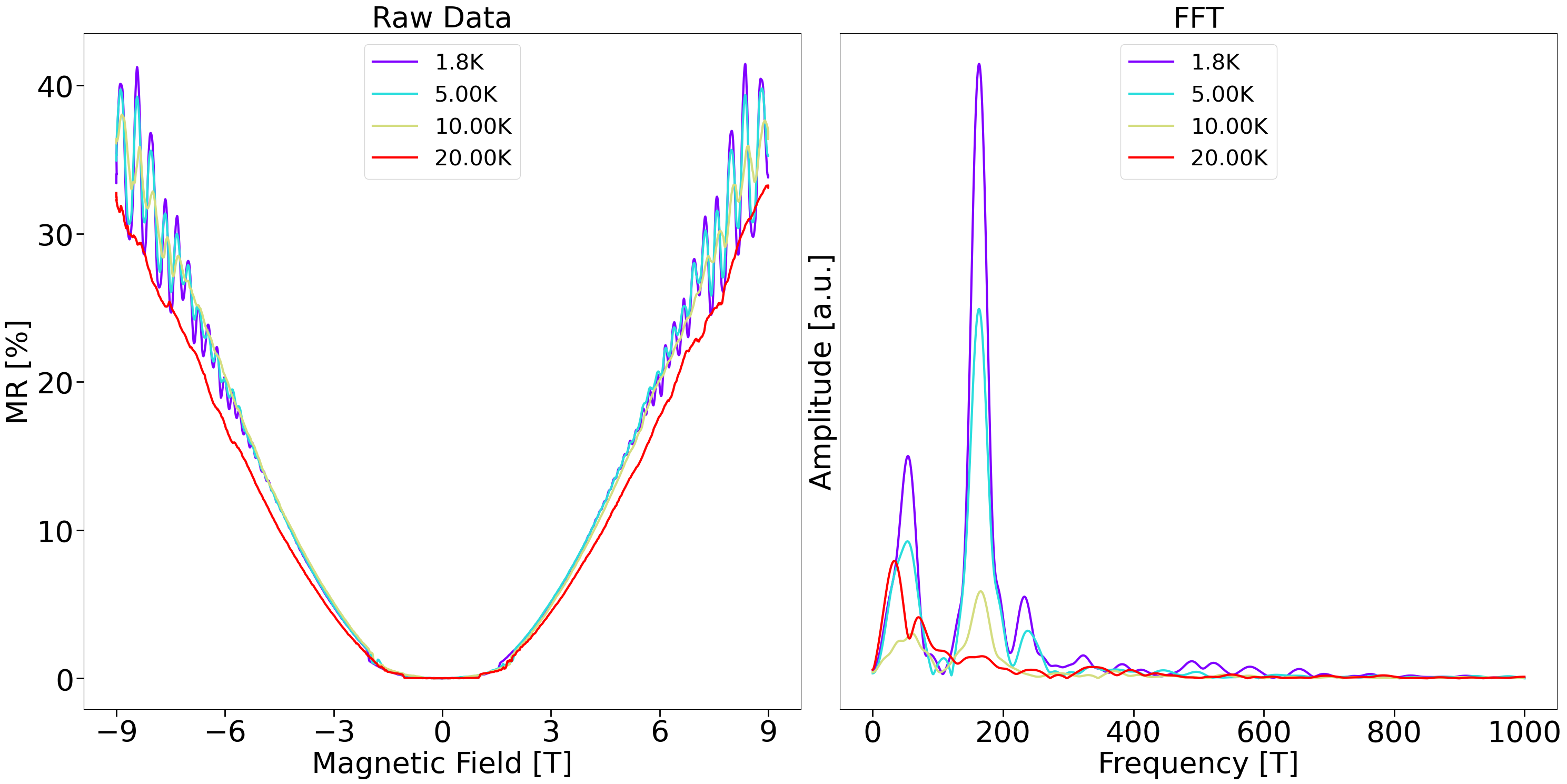}
  \caption[Magnetoresistance of a TaNiTe$_5$ sample up to \SI{9}{\tesla}.]{Left: Magnetoresistance of a TaNiTe$_5$ sample at various temperatures. Total magnetoresistance derived from the in-phase (IP) and out-of-phase (OOP) resistances via ${R}=\sqrt{{R}_{\text{IP}}^2+{R}_{\text{OOP}}^2}$ is shown here. The frequency on the lock-in amplifier was set to \SI{11}{\hertz}. Right: Respective fast Fourier transforms.}
  \label{TaNiTe5_magnetoresistance_overview}
\end{figure}

Generally, a lock-in amplifier applies a periodic current with a frequency and phase shift that it can use to amplify the measured voltage drop across the sample. The measured signal can be separated into an oscillation that is in phase with the reference signal, and one that is shifted by $\pi/2$ with respect to the reference signal of the lock-in amplifier, and is thus out of phase. We should expect the vast majority of the input signal to be in phase with the reference signal unless inductive or capacitive effects are present. \autoref{TaNiTe5_magic_osc} shows an enhanced signal in these components both of which host strongly amplified quantum oscillations, with the out-of-phase component even obtaining negative values at high magnetic fields. At fields between \SI{-9}{\tesla} to \SI{9}{\tesla} the resistance in the in-phase component shows semiclassical behaviour, while the out-of-phase component increases quadratically in field. Both components show a drop at fields beyond $\pm\SI{9}{\tesla}$ with continuously increasing quantum oscillations. To investigate the origin of this peculiar effect, the measurements were repeated in the \SI{9}{\tesla}-PPMS with the same sample, again with $I\parallel a$ and $B\parallel b$, and the temperature-dependent magnetoresistance defined in terms of longitudinal resistance $R(B)$
\begin{equation}\label{semiclassical_MR}
	\text{MR} = \frac{R(B)-R(0)}{R(0)}
\end{equation}
is illustrated in \autoref{TaNiTe5_magnetoresistance_overview}. Higher resolution was obtained in this measurement which allowed for the analysis of quantum oscillations. We find frequencies of \SI{53.0(3)}{\tesla}, \SI{163.0(4)}{\tesla} and \SI{233(3)}{\tesla} with effective masses of \SI{0.14(2)}{\electronmass}, \SI{0.23(1)}{\electronmass} and \SI{0.23(3)}{\electronmass}, respectively. The Dingle temperature for the first two frequencies could be determined as \SI{12.9(4)}{\kelvin} and \SI{14.7(3)}{\kelvin} respectively, while the data for the third frequency did not allow for determination of the Dingle temperature. Overall agreement with the literature \cite{daschner2024probing,xu2020anisotropic,chen2021three,ye2022anisotropic} is good and this data provides evidence of Shubnikov - de Haas oscillations, which so far have only been reported at  much higher fields of up to \SI{30}{\tesla} \cite{chen2021three} or even up to \SI{50}{\tesla} \cite{xu2020anisotropic}.

Given that the frequencies and effective masses are consistent with DFT calculations and results from previous measurements \cite{daschner2024probing}, these strongly amplified oscillations seem to fundamentally originate from the Fermi surface of this material. The purity of the crystals used here can be ruled out as a potential cause for this enhancement as the residual resistance ratio (RRR) is close to the value in the literature (RRR$\approx25$). Even with a high RRR, the purity of the crystals alone is not enough to explain the large out-of-phase component. Spot-welded gold wires and gold wires attached to the sample with silver epoxy have both shown amplified oscillations, indicating that this effect is not a result of the wiring. In particular any chemical reaction between the silver epoxy and the sample surface does not contribute to these oscillations.

\begin{figure}[t]
  \centering
  \includegraphics[width=\columnwidth]{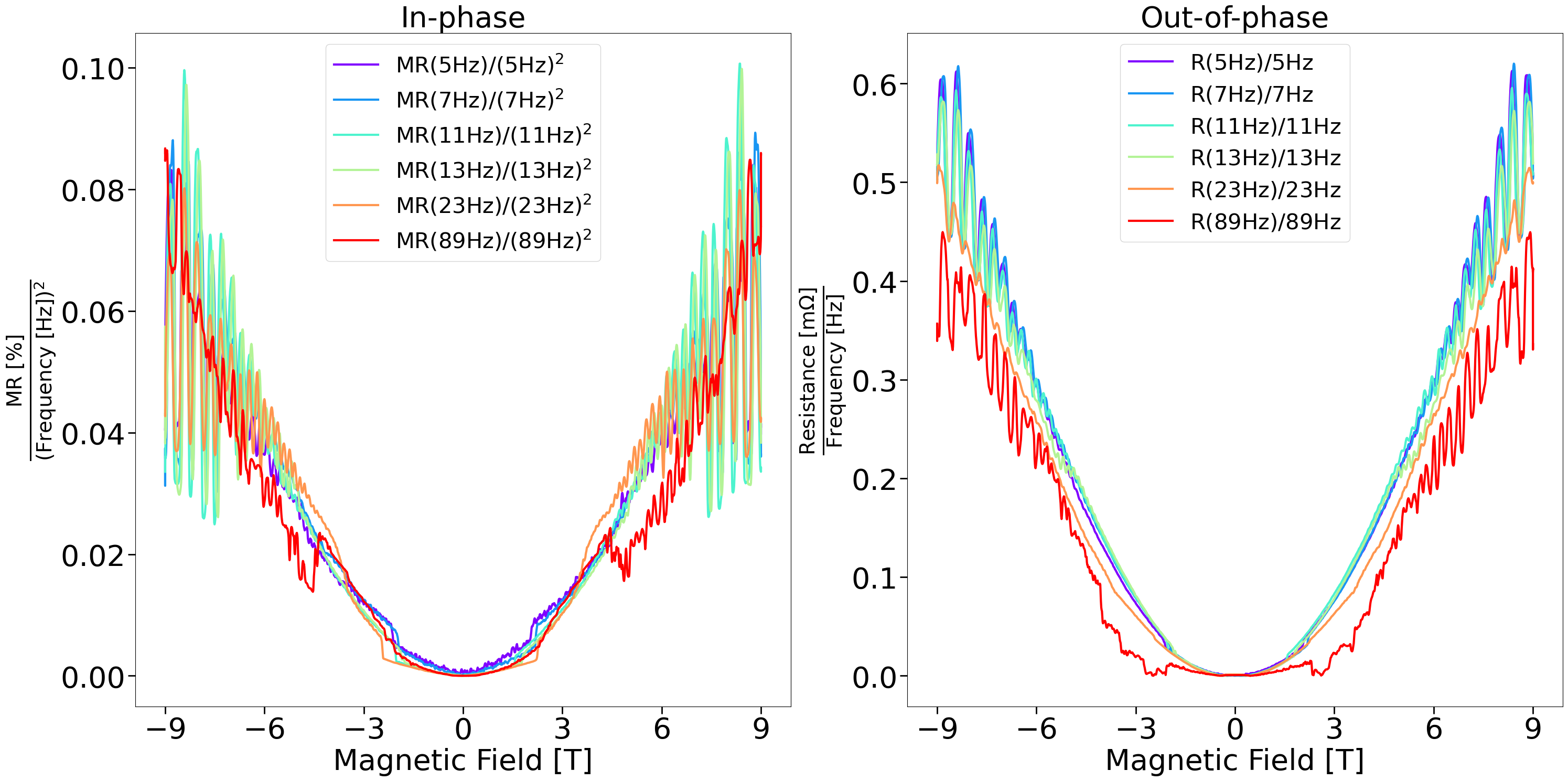}
  \caption[Data from \autoref{TaNiTe5_magnetoresistance_overview} split into the in-phase and out-of-phase components.]{Data from \autoref{TaNiTe5_magnetoresistance_overview} measured at \SI{1.8}{\kelvin} split into the in-phase and out-of-phase component of the lock-in amplifier. Since the out-of phase component is zero at \SI{0}{\tesla}, the resistance is shown instead of the MR in this case.}
  \label{TaNiTe5_magnetoresistance_X-Y_freq}
\end{figure}

To perform a more thorough analysis of the data shown in \autoref{TaNiTe5_magnetoresistance_overview}, the low-temperature data was split into its in-phase and out-of-phase components similar to the data obtained in the \SI{14}{\tesla}-PPMS as illustrated in \autoref{TaNiTe5_magnetoresistance_X-Y_freq}. Note that for the out-of-phase component the resistance at $B=0$ is zero, and since the magnetoresistance is only well-defined if $R(0)\neq 0$, the resistance $R(B)$ is given instead of MR in this case. Apart from its quadratic dependence on the magnetic field, the magnetoresistance of the in-phase component also shows a quadratic dependence on the frequency of the reference signal set on the lock-in amplifier, which is evident when dividing the data by the square of that frequency. The resistance of the out-of-phase component is also quadratic in $B$, but shows a linear dependence on the frequency of the reference signal. At higher frequencies ($\sim\SI{89}{\hertz}$) this dependence seems to diverge from the linear relationship, indicating that the linearity only holds to first order.

\begin{figure}[t]
  \centering
  \includegraphics[width=\columnwidth]{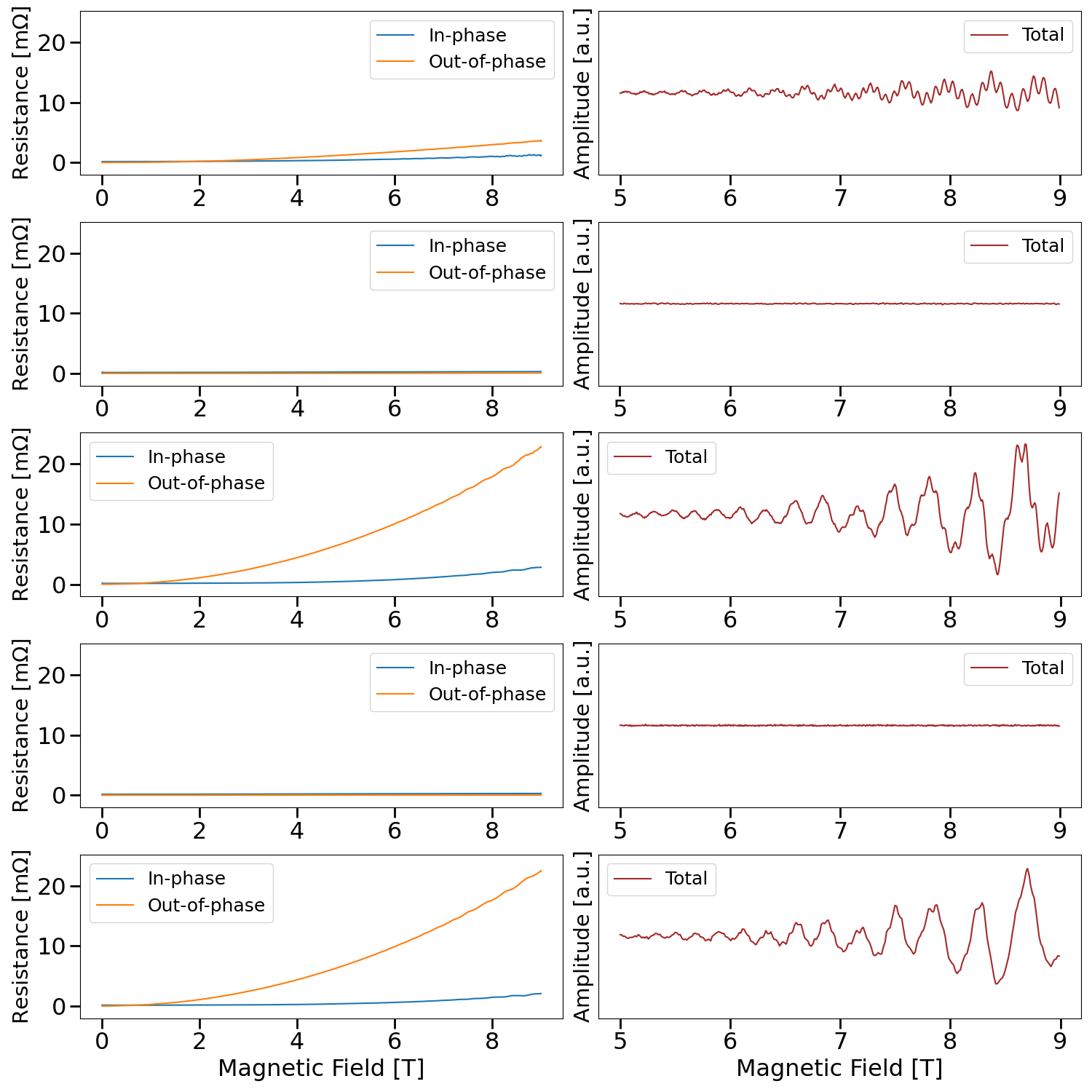}
  \caption[Relation between enhancement of quantum oscillations and mechanical movement.]{Relation between enhancement of quantum oscillations and mechanical movement. Left: Resistance in field of a TaNiTe$_5$ sample that is either solely held up by thin gold wires (first, third and fifth row) or attached to the measurement puck using vacuum grease (second and fourth row) measured at \SI{1.8}{\kelvin}. Directions of current and magnetic field are $I\parallel a$ and $B\parallel b$. Right: Background subtracted data, where the total data shown here includes the in-phase and out-of-phase components. The scale is identical in all figures, i.e. despite being arbitrary, the units are the same in all figures.}
  \label{glued_notglued_TaNiTe5_MR_lockin1}
\end{figure}

Motivated by the occurrence of mechanical signatures such as resonance peaks in measurements of TaNiTe$_5$ and other materials even at room temperature (see \appref{mechanical_Fe3GeTe2}), the quantum oscillation data shown here can possibly be explained by the interplay of classical and quantum effects. To verify this hypothesis, samples of TaNiTe$_5$ were measured such that mechanical movement would either be allowed or restricted. A series of measurements started with a free floating sample which is only held by the gold wires attached to it with the current and magnetic field again as sketched in \autoref{TaNiTe5_measurement_sketch_1}. After the first field sweep, the sample is taken out of the \SI{9}{\tesla}-PPMS and firmly attached to the bottom of the measurement puck with vacuum grease before being cooled down and measured in another field sweep. The sample is then again taken out of the cryostat, detached from the vacuum grease so that it is only held up by the gold wires, and the process repeats. Field sweeps with samples that are floating freely show strong quantum oscillations as illustrated in \autoref{glued_notglued_TaNiTe5_MR_lockin1}, while field sweeps with samples that are attached to the puck with vacuum grease do not show any oscillations at all. Furthermore, a strong out-of-phase component arises only in field sweeps with detached samples and obtains values much larger than the in-phase component which leads to the conclusion that this strong out-of-phase component and the presence of quantum oscillations are related in some way.

\section{Theoretical Model}

To explain this strange behaviour, a simple model based on damped harmonic oscillators as motivated in \appref{mechanical_Fe3GeTe2} can be used. \autoref{TaNiTe5_measurement_sketch_1} shows a regular setup of a four-point measurement with the gold wires simultaneously measuring and holding up a sample of length L, where we can neglect gravity due to the relatively low weight of the sample compared to the spring constant of the wires. We thus have a floating sample through which an AC current $I(t)$ proportional to the voltage
\begin{equation}
	V(t) = V_0 \sin{(\omega t)}
\end{equation}
is flowing inside an external magnetic field. Here $V_0$ is the amplitude and $\omega$ is the frequency of the signal applied by the lock-in amplifier. The sample can be approximated by a straight, sturdy, one-dimensional rod whose length is parametrised by
\begin{equation}
	\mathbf{L} = \text{L}\ \mathbf{\hat{L}} = \text{L}\
	\begin{pmatrix}
		\cos{(\theta_0+\theta(t))}\\
		\sin{(\theta_0+\theta(t))}\\
		0
	\end{pmatrix}
\end{equation}
while the magnetic field is given by $\mathbf{B} = \text{B}\mathbf{\hat{y}}$.

Since a current is flowing through the sample inside a magnetic field, it is affected by the Lorentz force
\begin{align}\label{Lorentz_force}
	\begin{split}
	\mathbf{F_L} &= I(t)\mathbf{L}\times\mathbf{B} = I(t)B\text{L}\ \mathbf{\hat{L}}\times\mathbf{\hat{y}} \\
		     &= I(t)B\text{L}\ \cos{\left(\theta_0+\theta(t)\right)}\ \mathbf{\hat{z}}
	\end{split}
\end{align}
that pushes the sample either in the +$\hat{\mathbf{z}}$ or -$\hat{\mathbf{z}}$ direction depending on the direction of the current. The current is along $\hat{\mathbf{x}}$ unless the sample rotates in space to an angle $\theta_0+\theta(t)$ which would reduce the Lorentz force due to the cosine term. We assume that the sample is initially tilted by a small angle $\theta_0$, and can rotate further by an angle of $\theta(t)$. Sources that induce such rotation will be discussed later.

\begin{figure}[t]
  \centering
  \includegraphics[width=0.8\columnwidth]{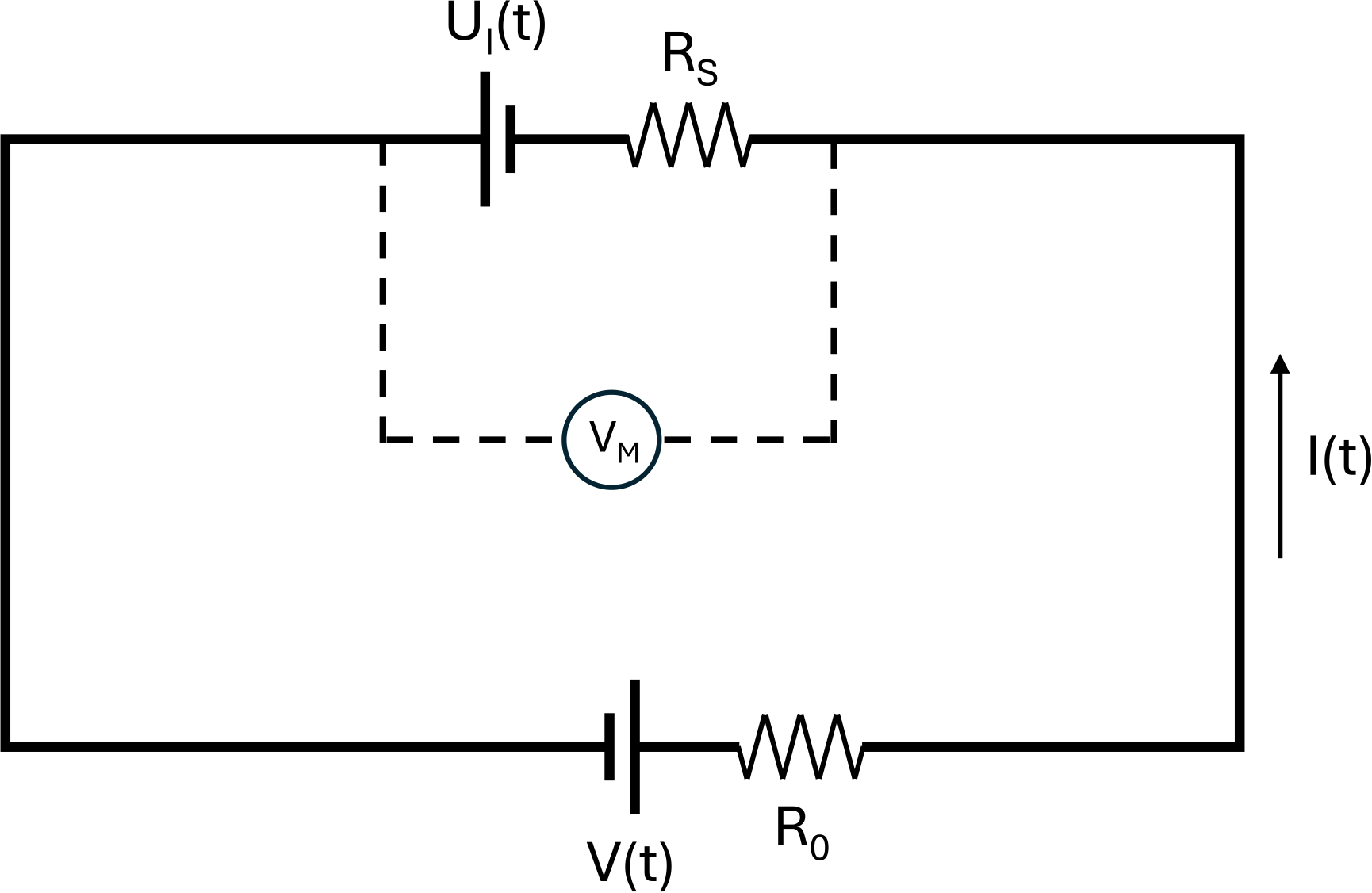}
  \caption[Simple circuit diagram for a four-point measurement including an additional induced voltage $U_I(t)$.]{Simple circuit diagram for a four-point measurement of a sample with resistance $R_S$ including an additional induced voltage $U_I(t)$. A lock-in amplifier generates a sinusoidal voltage $V(t)$ and measures a voltage $V_M$. The total resistance of the circuit includes a resistor $R_0=\SI{1}{\kilo\ohm}$ in series that was put in place to reduce the current, the resistance of the wires, and the resistance of the sample.}
  \label{circuit_TNT_magic}
\end{figure}

As a result of the Lorentz force, the sample is expected to oscillate back and forth at the frequency of the AC current, where the spring constant of the gold wires act as a damping term in the equation of motion (e.o.m.) which we will set up shortly. Now, whenever a metallic sample is moving in a magnetic field, independent of the exact cause of such movement, we should also expect an induced voltage
{\small
\begin{align}\label{induced_voltage}
	\begin{split}
		U_I(t) &= (\mathbf{\dot{r}}(t)\times\mathbf{B})\cdot \mathbf{L} = -\dot{z}(t)B\text{L}\ \mathbf{\hat{x}}\cdot \mathbf{\hat{L}} \\
		       &= -\dot{z}(t)B\text{L}\ \cos\left(\theta_0+\theta(t)\right)
	\end{split}
\end{align}
}
inside the sample, which is proportional to the velocity along $\hat{\textbf{z}}$. This induced voltage will be measured by the voltage contacts together with the magnetoresistance of the sample as sketched in \autoref{circuit_TNT_magic}. \autoref{Lorentz_force} and \autoref{induced_voltage} indicate that the amplitude of both, the induced voltage and the Lorentz force, is strongly dependent on the angle between the current $I$ and the magnetic field $B$. If a torque was present that tilts the orientation of the sample by $\theta(t)$, while the sample is constantly oscillating back and forth due to the AC current, both the Lorentz force and the induced voltage would be lowered. A potential candidate for such a torque is given by de Haas - van Alphen oscillations in the magnetic torque that can indeed cause a sample to tilt in field. Such torque  $\tau_{QO}$ is just given \cite{shoenberg1962fermi} as

\begin{equation}\label{torque_LK}
	\resizebox{0.9\linewidth}{!}{$\tau_{QO}\left(\theta,B\right) \propto B^{\frac{3}{2}} R_T R_D R_S \sin\left(2\pi \left(\frac{F\left(\theta\right)}{B}-\gamma\right)\pm\delta\right)$}
\end{equation}
if only one orbit and one harmonic are taken into account. Here, $R_x$ (x=T,D,S) are various damping terms, while $F\left(\theta\right)$ is the frequency arising from the respective Fermi surface and $B$ the magnetic field. $\gamma$ and $\delta$ are shifts associated with the Berry phase and the dimensionality of the system and play very little role in the model considered here and will thus be neglected. Note that in case the sample made a \SI{180}{\degree} turn, \autoref{Lorentz_force} and \autoref{induced_voltage} remain correct, however the contacts of the gold wires flip as well, and hence the resulting sign change needs to be incorporated in this model. This can be done by taking the absolute value of the cosine in \autoref{induced_voltage}.

The underlying principle of this model is that the measurements of quantum oscillations in the resistivity, also known as Shubnikov - de Haas oscillations, shown here are in reality measurements of the induced voltage through the constant back and forth movement of the sample along $\hat{\mathbf{z}}$ and its reduction or increase due to the rotation of the sample in the $\hat{\textbf{x}}$-$\hat{\textbf{y}}$ plane caused by de Haas - van Alphen oscillations in the magnetic torque.
Given this qualitative understanding, we can attempt to continue with the model by motivating two equations of motion for the movement and tilt of the sample. We introduced $z(t)$ as the only degree of freedom that describes the movement of the centre of mass, since movements in the $\hat{\mathbf{x}}$ and $\hat{\mathbf{y}}$ direction do not contribute to the induced voltage. Furthermore, we introduced the angle $\theta(t)$ which describes the orientation of the sample in the $\hat{\mathbf{x}}$-$\hat{\mathbf{y}}$ plane. In what follows, we will motivate equations of motion for each of these two degrees of freedom, both of which will be based on damped harmonic oscillators. They will then be coupled which allows us to derive an expression for the overall induced voltage in the sample. We begin with the angle $\theta(t)$, since it can be solved independently of the lateral movement along $\hat{\mathbf{z}}$. Once an expression for $\theta(t)$ is derived, it can be used to find $z(t)$ which can then be used to find an explicit expression for $U_I(t)$ in \autoref{induced_voltage}. Together with the sample's inherent magnetoresistance $R_S$, we propose that this is what is measured in the empirical data shown here.

The angle $\theta(t)$ is only affected by the magnetic torque arising from quantum oscillations, which are damped due to the presence of gold wires attached to the sample (see \autoref{TaNiTe5_measurement_sketch_1}) acting as springs. It represents the orientation of the sample and can be described by a damped harmonic oscillator:
\begin{equation}\label{eom_theta}
	J \ddot{\theta}(t) = \tau_{QO}\left(\theta(t),t\right)-k_1 \theta(t)-k_2 \dot{\theta}(t)
\end{equation}
where $t$ is the time, $J$ is the moment of inertia, $k_1$ and $k_2$ are damping constants.
Given that the magnetic field is ramped up comparatively slowly ($B_{\text{rate}}=\sim$\SI{0.1}{\tesla\per\minute}), we can neglect the first and second derivatives such that \autoref{eom_theta} becomes
\begin{equation}\label{eom_theta_tilde}
	\theta(t) \approx \theta = \frac{\tau_{QO}\left(\theta,B\right)}{k_1}
\end{equation}
where the magnetic field $B$ can be related to time through $B = B_{\text{rate}}\cdot t$. At any given field, we thus assume that the orientation of the sample will instantaneously align at an angle which can be found using the equation above. Note that since $\tau_{QO}\left(\theta,B\right)$ also depends on the angle, which it does in general since the frequency is angle-dependent, solving the above equation can become complicated. Nonetheless, the problem of solving a differential equation was reduced to solving an algebraic equation which is in general much simpler. For reasons of numerical stability, we will assume that the frequency $F$ is angle-independent from here on.

\begin{figure}[t]
     \includegraphics[width=0.48\columnwidth]{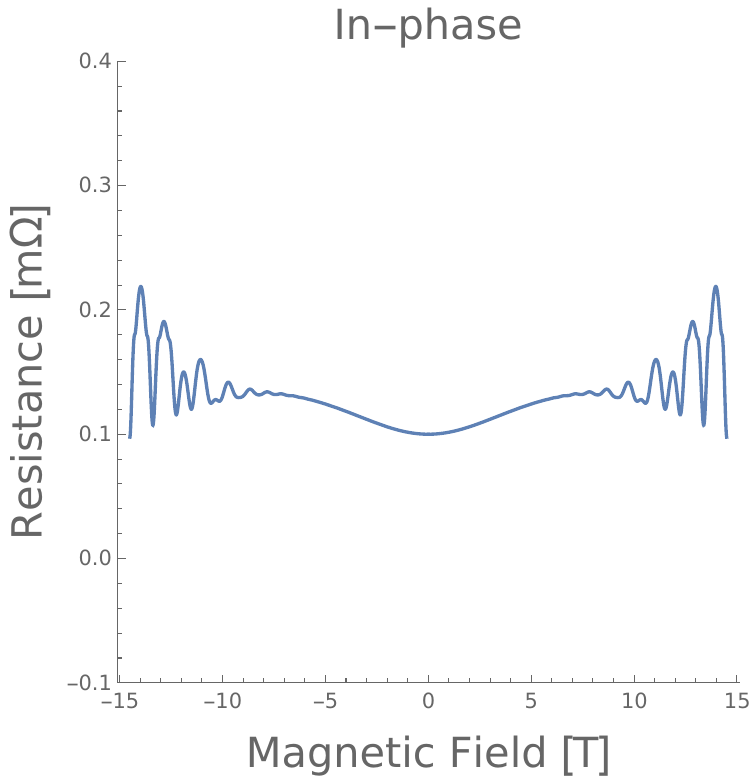}
     \includegraphics[width=0.48\columnwidth]{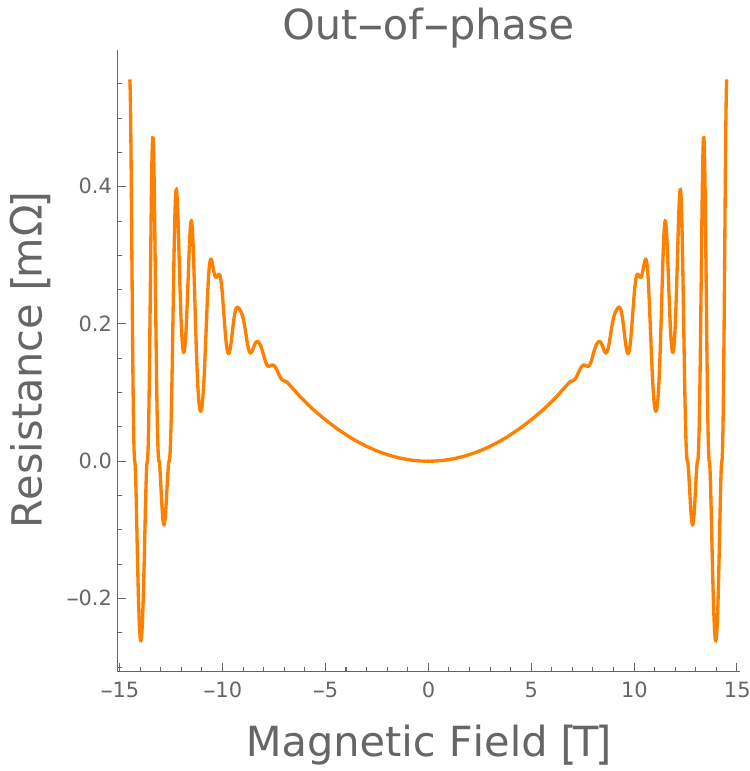}
     \caption[Numerical evaluation of \autoref{magic_x-component} and \autoref{magic_y-component}.]{Numerical evaluation of \autoref{magic_x-component} and \autoref{magic_y-component} closely resembling \autoref{TaNiTe5_magic_osc}.}\label{TaNiTe5_mathematica_magic}
\end{figure}

The equation of motion for the centre of mass movement along $\hat{\mathbf{z}}$ can be described by a driven, damped harmonic oscillator involving the Lorentz force $\mathbf{F_L}$ and damping terms $\kappa_1$ and $\kappa_2$ arising from the gold wires:
{\small
\begin{align}
	m \mathbf{\ddot{r}}(t) &= \mathbf{F_L} - \kappa_1 \mathbf{r}(t) - \kappa_2 \mathbf{\dot{r}}(t)\\
	m \ddot{z}(t)\ \mathbf{\hat{z}} &= \left[I(t)B\text{L}\ \cos{\left(\theta_0+\theta\right)} - \kappa_1 z(t) - \kappa_2 \dot{z}(t)\right]\ \mathbf{\hat{z}} \label{eom_r}
\end{align}
}
where the centre of mass of the sample is only moving along the $\mathbf{\hat{z}}$ direction. The periodic Lorentz force is driven by the periodic current from the lock-in amplifier. In principle, the current $I(t)$ itself also needs to be modified since the movement of the sample will induce a voltage that does not only affect the measurement, but also the current flowing through the circuit. This would lead to a set of coupled equations of motion, since the modified current $I(t)$ in turn also impacts the movement of the sample through \autoref{eom_r}. However, we can safely assume that the current $I(t)$ is dominated by the lock-in amplifier, i.e. $V(t)\gg U_I(t)$, hence the current can be expressed by
\begin{equation}\label{current_TNT_magic}
	I(t) = \frac{1}{R_0}(V(t)+U_I(t)) \approx \frac{1}{R_0}V(t) = \frac{V_0}{R_0} \sin{(\omega t)}
\end{equation}

Since the resistance of the wires and of the sample are orders of magnitude smaller than $R_0$ (see \autoref{circuit_TNT_magic}), they will both be neglected here. The goal is to solve the e.o.m. for $z(t)$ in \autoref{eom_r} to obtain $U_I(t)$ in \autoref{induced_voltage}. The measured voltage $V_M$ is then given by
\begin{equation}\label{voltage_measured_magic}
	V_M(t) =  I(t) R_{S}(B) + U_I(t)
\end{equation}
which is just a modification of the sample resistance with the induced voltage. The sample resistance $R_S(B)$ in field is assumed to show semiclassical behaviour which can simply be expressed \cite{pippard1989magnetoresistance,ali2014large,fallah2016temperature,wang2015origin,pallecchi2011magnetotransport,wang2012multiband} as
\begin{equation}\label{semiclassical_R}
	R_S(B)  = R_S(0) + R_S(0) \frac{\alpha B^2}{\beta+B^2}
\end{equation}
where $\alpha$ and $\beta$ are material-dependent parameters that can be determined from a stationary sample. Note that it will be the induced voltage $U_I(t)$ that introduces the pronounced quantum oscillations both in the in-phase and out-of-phase component measured by the lock-in amplifier. After solving the second order differential equation in \autoref{eom_r} and using it in $V_M(t)$, the result can then be grouped by terms proportional to $\sin(\omega t)$ and $\cos(\omega t)$ to obtain the in-phase (IP) and out-of-phase (OOP) components, which leads to
{\small
\begin{align}
	V_M^{\text{IP}} &= \frac{V_0}{R_0}\left(R_S-\frac{\kappa_2 \omega^2 B^2 l^2 \cos\left(\theta_0+\theta\right)\left|\cos\left(\theta_0+\theta\right)\right|}{\kappa_1^2+m^2 \omega^4-2 \kappa_1 m \omega^2+\kappa_2^2 \omega^2}\right) \label{magic_x-component} \\
	V_M^{\text{OOP}} &= \frac{V_0}{R_0}\frac{\left(m \omega^2-\kappa_1\right) \omega B^2 l^2 \cos\left(\theta_0+\theta\right)\left|\cos\left(\theta_0+\theta\right)\right| }{\kappa_1^2+m^2 \omega^4-2 \kappa_1 m \omega^2+\kappa_2^2 \omega^2} \label{magic_y-component}
\end{align}
}
with the corresponding results being illustrated in \autoref{TaNiTe5_mathematica_magic}, showing strong resemblance with the data in \autoref{TaNiTe5_magic_osc}.
\begin{figure}[t]
  \centering
  \includegraphics[width=\columnwidth]{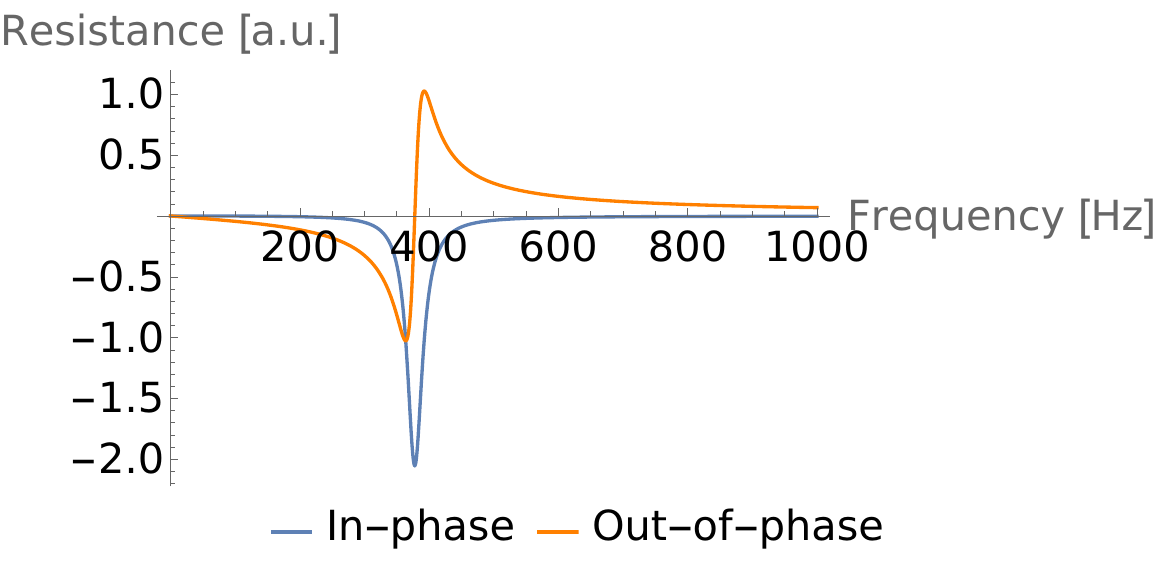}
  \caption[Mechanical resonance measured in the resistance described by \autoref{magic_x-component} and \autoref{magic_y-component}.]{Mechanical resonance measured in the resistance described by \autoref{magic_x-component} and \autoref{magic_y-component}. The resonant frequency can be determined using \autoref{resonant_frequency_magic}.}
  \label{TaNiTe5_mathematica_resonance}
\end{figure}
To compare this model to the measurement data further, we can generate a power series expansion of \autoref{magic_x-component} and \autoref{magic_y-component} to lowest order in the frequency $\omega$:

{\tiny
\begin{align}
	V_M^{\text{IP}} &\approx R_S\frac{V_0}{R_0}-\frac{V_0}{R_0}\frac{\kappa_2 l^2}{\kappa_1^2}\ \cos\left(\theta_0+\frac{\tau_{QO}}{k_1}\right) \left| \cos \left(\theta_0+\frac{\tau_{QO}}{k_1}\right)\right| B^2 \omega^2 \\
	V_M^{\text{OOP}} &\approx -\frac{V_0}{R_0}\frac{l^2}{\kappa_1}\  \cos \left(\theta_0+\frac{\tau_{QO}}{k_1}\right) \left| \cos \left(\theta_0+\frac{\tau_{QO}}{k_1}\right)\right|B^2 \omega
\end{align}
}
which confirms the quadratic and linear frequency dependences measured in the in-phase and out-of-phase components in \autoref{TaNiTe5_magnetoresistance_X-Y_freq}, respectively. The slight deviation from these power laws at higher frequencies in the measurement data could either be due to higher order terms or due to non-linear effects in the gold wires. Furthermore, we can see that the dependence on the magnetic field is quadratic, which was also observed in the raw data.

The resonance frequency $\omega_R$ can be found by determining the zeros of either \autoref{magic_y-component} or the derivative of \autoref{magic_x-component}, each of which leads to
\begin{equation}\label{resonant_frequency_magic}
	\omega_R = \sqrt{\frac{\kappa_1}{m}}
\end{equation}
This resonance can be reproduced in the in-phase and out-of-phase components as illustrated in \autoref{TaNiTe5_mathematica_resonance} at a fixed magnetic field. It strongly resembles the resonance peaks observed in \autoref{Fe3GeTe2_FieldSweep+TaNiTe5_resonance_peaks} b) up to a sign.

\section{Conclusion}

We discuss the occurrence of a novel mechanical effect that amplifies the magnitude of quantum oscillations. A simple model based on a driven, damped harmonic oscillator, with the applied current acting as the driving force and the gold wires introducing damping, is able to explain the results accurately, and no topological properties need to be invoked. The model presented here is consistent with a number of properties measured in samples of TaNiTe$_5$. Both the in-phase and out-of-phase magnetoresistances show quadratic behaviour in field, and depend quadratically and linearly on the frequency of the reference signal from the lock-in amplifier, respectively.

Future experimental studies of Dirac semimetals, and possibly metals more generally, should take these results into account when performing Shubnikov - de Haas measurements.

\begin{acknowledgments}
M. Daschner would like to thank Jiasheng Chen for many discussions regarding the nature of these strong quantum oscillations.
\end{acknowledgments}

\appendix

\section{Image of a real setup in the \SI{9}{\tesla}-PPMS}\label{TaNiTe5_on_puck}

\begin{figure}[ht]
  \includegraphics[width=0.48\columnwidth]{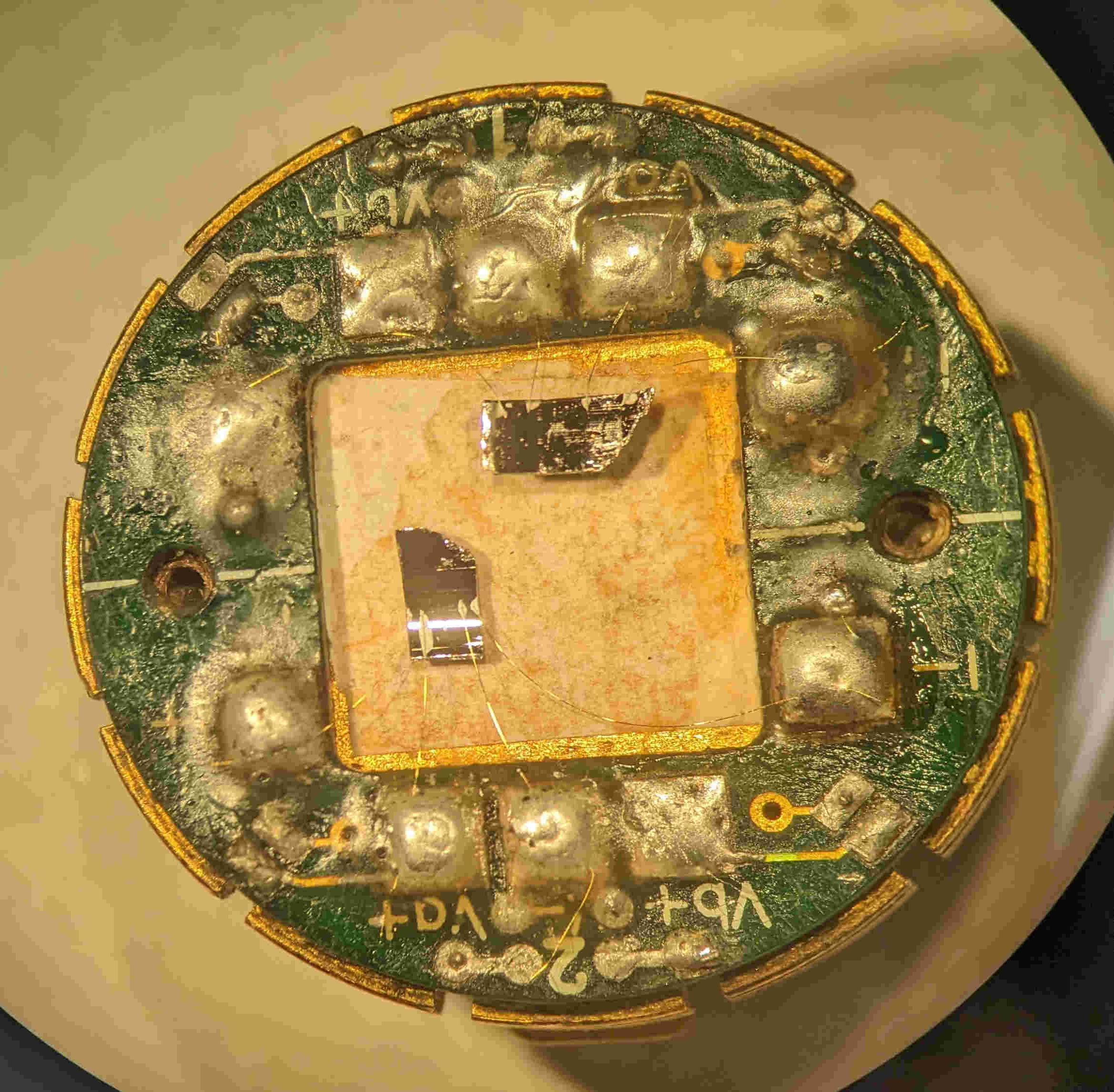}
  \includegraphics[width=0.48\columnwidth]{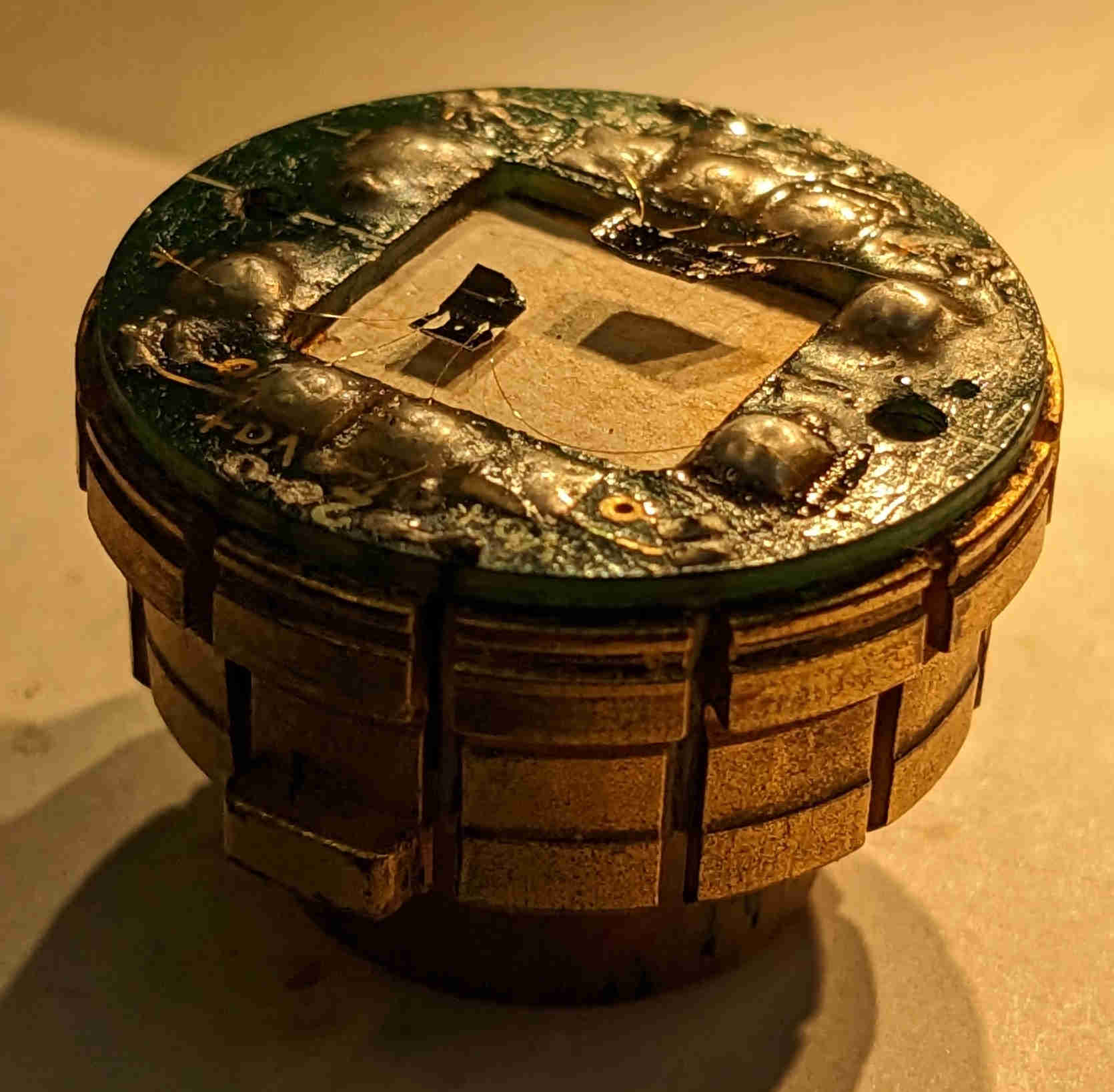}
  \caption[Two TaNiTe$_5$ samples on a standard Quantum Design measurement puck.]{Two TaNiTe$_5$ samples on a standard Quantum Design measurement puck. Note how both of them are held by the gold wires alone, thus floating above the platform.}
\end{figure}

\section{Mechanical resonance in magnetic field}\label{mechanical_Fe3GeTe2}

\begin{figure}[ht]
	\includegraphics[width=0.48\columnwidth]{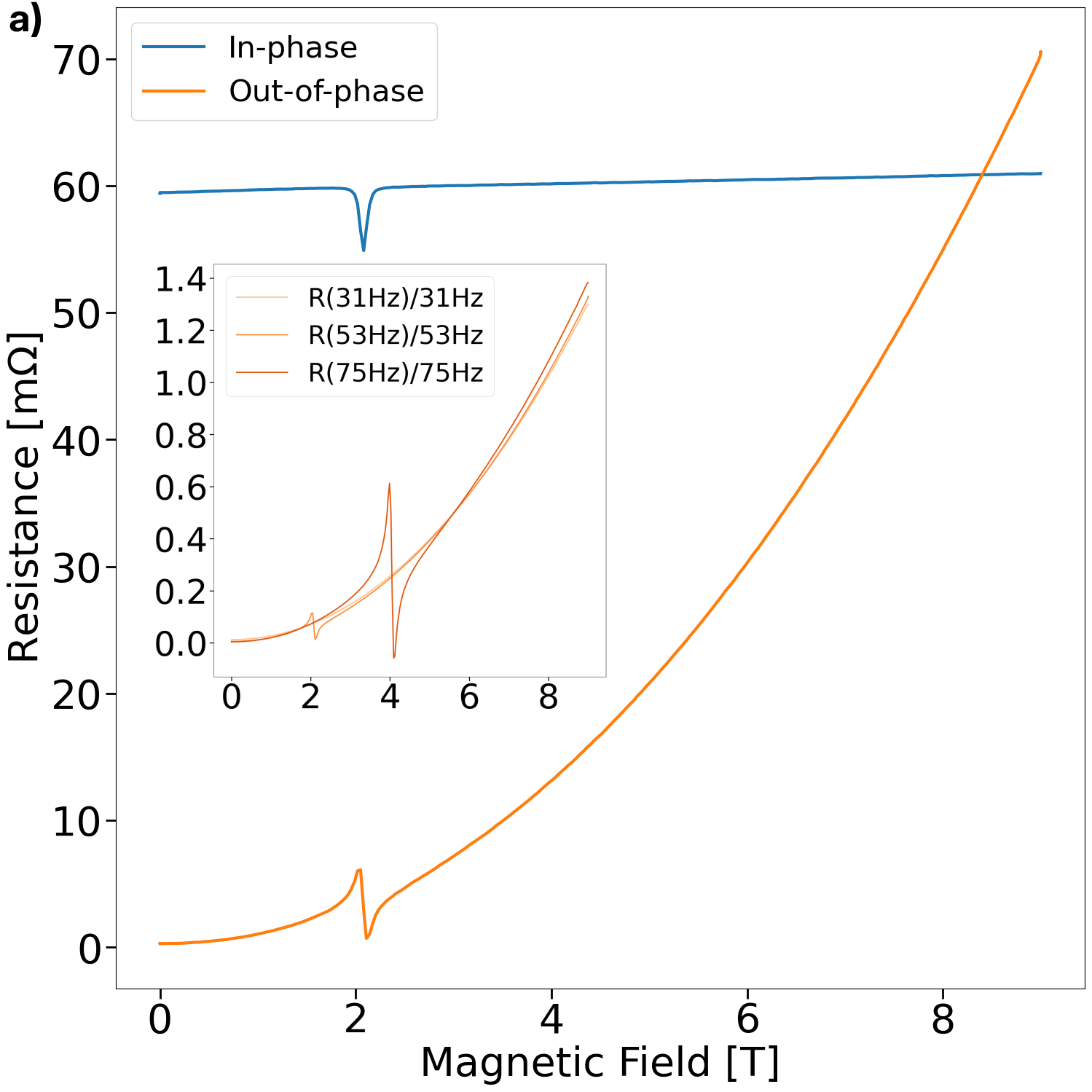}
	\includegraphics[width=0.48\columnwidth]{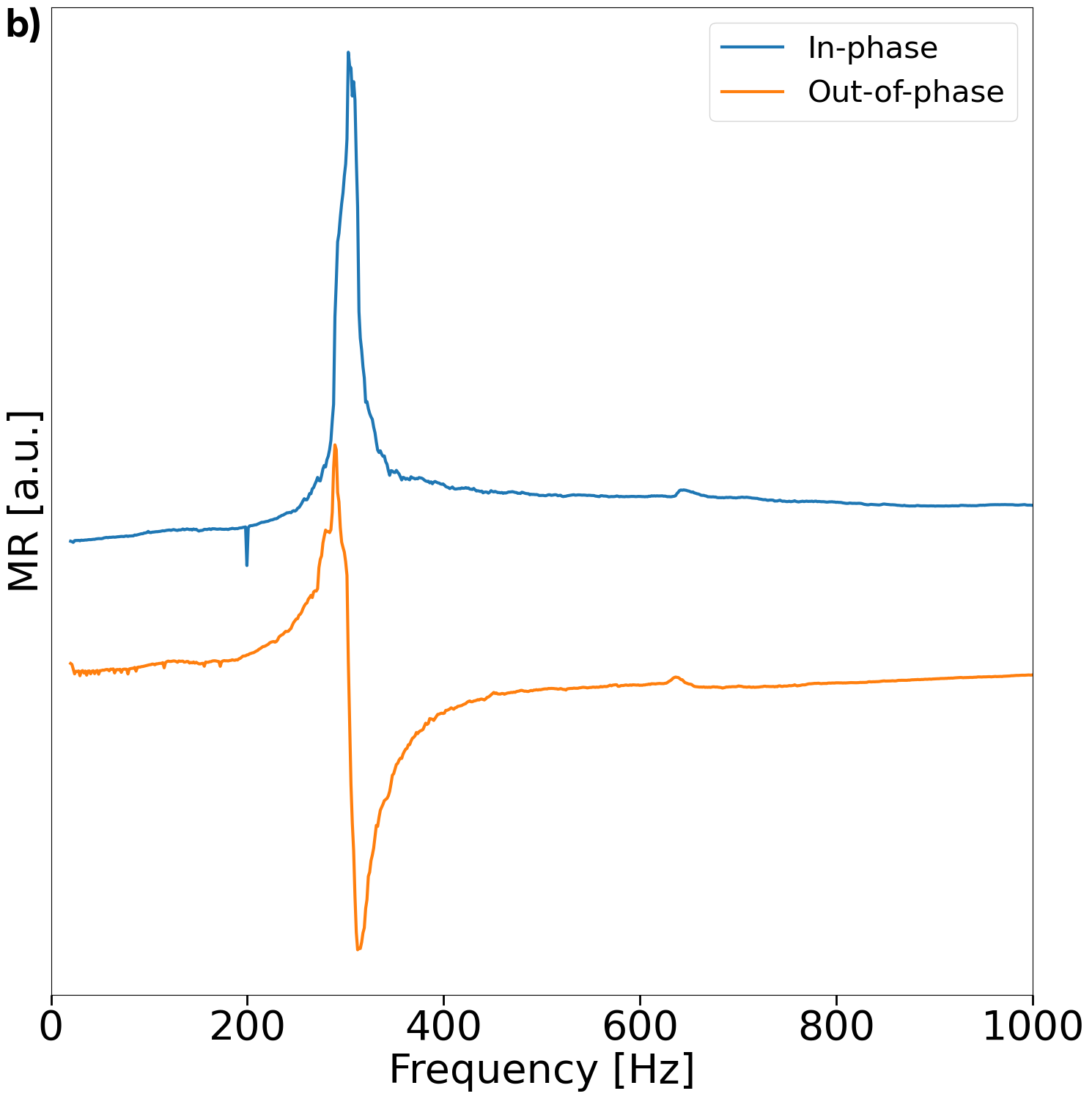}
	\caption{a) Field sweep of a Fe$_3$GeTe$_2$ sample at \SI{53}{\hertz} and \SI{300.0}{\kelvin}. The inset shows the out-of-phase component for different frequencies divided by the respective frequency. b) Frequency sweep of a TaNiTe$_5$ at \SI{13}{\tesla} and \SI{1.5}{\kelvin}. Both components resemble the behaviour of a damped harmonic oscillator known from classical mechanics.}\label{Fe3GeTe2_FieldSweep+TaNiTe5_resonance_peaks}
\end{figure}

The strong out-of-phase component with its linear frequency-dependence in the main text was also observed in other materials whenever the respective samples had not been firmly attached to the measurement platform. This effect was present even at room temperature measured in the \SI{9}{\tesla}-PPMS as is illustrated with a sample of Fe$_3$GeTe$_2$ shown in \autoref{Fe3GeTe2_FieldSweep+TaNiTe5_resonance_peaks} a). While the in-phase component increases only slightly in magnetic field, which is expected for such a measurement at room temperature, the out-of-phase component increases $\propto B^{\alpha}$ with $\alpha=2.091(2)$ and thus nearly quadratically in field. Furthermore, the inset clearly shows a linear dependence on the frequency of the reference signal. Interestingly, this data also shows the presence of a resonance peak emerging at around \SI{2}{\tesla} both in the in-phase and out-of-phase components, strongly resembling the physics of a damped harmonic oscillator. The position of this resonance peak seems to shift as the frequency is altered which is illustrated in the inset in \autoref{Fe3GeTe2_FieldSweep+TaNiTe5_resonance_peaks} a). To see if these resonance peaks also appear in TaNiTe$_5$, a sample of this material was measured in a custom-designed Oxford Instruments cryostat at \SI{1.5}{\kelvin} in a constant external magnetic field of \SI{13}{\tesla}, and the frequency was slowly increased (\SI{8}{\hertz\per\minute}), with the results given in \autoref{Fe3GeTe2_FieldSweep+TaNiTe5_resonance_peaks} b). Upon increasing the frequency on the lock-in amplifier a clear resonance peak becomes visible, reminiscent of what was observed in Fe$_3$GeTe$_2$ when the magnetic field was swept at constant frequency. These results further confirm the presence of mechanical effects in these measurements.

\newpage

%\onecolumngrid

\bibliographystyle{unsrt}  % Choose your bibliography style
\bibliography{MAGIC_QO_TaNiTe5.bib}

\end{document}